\documentclass[twocolumn,showpacs,preprintnumbers,amsmath,amssymb,prb]{revtex4}

\usepackage{graphicx}
\usepackage{dcolumn}
\usepackage{bm}
\usepackage{amssymb}
\begin{document}

\title{Persistent current induced by quantum light}
\author{O. V. Kibis}\email{Oleg.Kibis@nstu.ru}

\affiliation{Department of Applied and Theoretical Physics,
Novosibirsk State Technical University, Karl Marx Avenue 20,
630092 Novosibirsk, Russia}


\begin{abstract}
It is demonstrated that the strong coupling of an electron gas to
photons in systems with broken time-reversal symmetry results in
bound electron-photon states which cannot be backscattered
elastically. As a consequence, the electron gas can flow without
dissipation. This quantum macroscopic phenomenon leads to the
unconventional superconductivity which is analyzed theoretically
for a two-dimensional electron system in a semiconductor quantum
well exposed to an in-plane magnetic field.
\end{abstract}


\maketitle

\section{Introduction} Advances in laser physics achieved in
recent decades have made possible the use of lasers as tools to
manipulate the electronic properties of various quantum systems.
Since the strong interaction between electrons and an intense
laser field cannot be described as a weak perturbation, it is
necessary to consider the system ``electron + field'' as a whole.
Such a bound electron-photon object, which was called ``electron
dressed by field'' (dressed electron), became commonly used model
in modern physics. \cite{Cohen-Tannoudji_b98,Scully_b01} The
field-induced modification of the energy spectrum of dressed
electrons
--- also known as a dynamic (ac) Stark effect --- was discovered
in both atoms \cite{Autler_55} and solids \cite{Mysyrowicz_86}
many years ago and has been studied in various electronic systems.
Particularly, it is well known that the interaction between a
solid and a monochromatic electormagnetic field can open energy
gaps $\Delta\varepsilon$ within electron energy bands of the solid
(see, e.g., Refs.
\onlinecite{Galitskii_70,Elesin_69,Galitskii_86,Vu_04}). Such a
gap opening is pictured schematically in Fig.~1(a) for the case of
interaction between a bulk semiconductor with the bandgap
$\varepsilon_g$ and an electromagnetic field with the frequency
$\omega_0>\varepsilon_g/\hbar$ (\onlinecite{Elesin_69}). It should
be stressed that the gaps $\Delta\varepsilon$ are opened in the
resonant points of $\mathbf{k}$--space (i.e., at electron wave
vectors $\mathbf{k}$ satisfying the condition of ``the photon
energy $\hbar\omega_0$ is equal to the energy interval between
electron bands''). If the electron energy spectrum of the solid is
symmetric, $\varepsilon(\mathbf{k})=\varepsilon(-\mathbf{k})$, the
resonant points --- and, correspondingly, the gaps
$\Delta\varepsilon$ --- are positioned symmetrically in the
$\mathbf{k}$--space with respect to band edges [see Fig.~1(a)].
Though the light-induced gap opening has been known for a long
time \cite{Galitskii_70,Elesin_69}, its theory is developed
exclusively for solids with such a symmetric electron energy
spectrum. Electronic systems with an asymmetric energy spectrum,
$\varepsilon(\mathbf{k})\neq\varepsilon(-\mathbf{k})$, escaped
attention before and will be considered below.
\begin{figure}[th]
\includegraphics[width=0.48\textwidth]{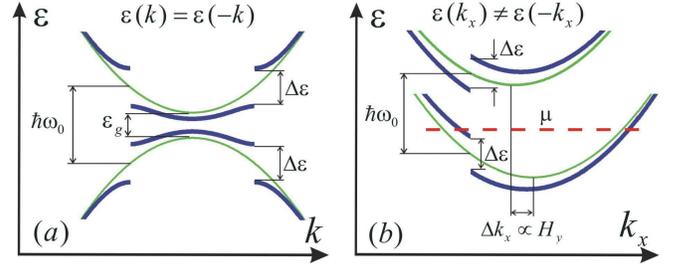}
\caption{(Color online) Energy spectrum of free electrons (thin
lines) and electrons dressed by an electromagnetic field with the
frequency $\omega_0$ (solid lines): (a) conductivity band and
valence band in a bulk semiconductor with the bandgap
$\varepsilon_g$; (b) first two electron subbands in an asymmetric
quantum well exposed to an in-plane magnetic field $H_y$. The
Fermi level of the electron system $\mu$ is pictured by the dashed
line.}
\end{figure}

It follows from the fundamentals of quantum mechanics that the
asymmetric energy spectrum of electrons can exist in systems with
broken both time-reversal symmetry and inversion symmetry.
Particularly, it takes place in nanostructures without an
inversion center in the presence of a magnetic field, including
asymmetric quantum wells
\cite{Kibis_97,Kibis_98_1,Kibis_98_2,Kibis_99,Kibis_00}, chiral
carbon nanotubes \cite{Kibis_93,Kibis_01_1,Kibis_01_2}, hybrid
semiconductor/ferromagnet nanostructures \cite{Lawton_02}, and so
on. For definiteness, let us consider such a simple nanostructure
as a quantum well (QW). Generally, a QW confines charge carriers,
which were originally free to move in three dimensions $(x,y,z)$,
to two dimensions $(x,y)$, forcing them to occupy a planar region
inside the QW. \cite{Ando_82} As a consequence, the physical
properties of a QW depend on the potential energy describing this
confinement, $U(z)$. In what follows, we will consider a QW with a
confining potential $U(z)$ devoid of an inversion center
(asymmetric QW). Technologically, such asymmetric QWs are
fabricated on the basis of semiconductor heterojunctions (see,
e.g., Refs.~\onlinecite{Ando_82,Nag_b00}). If an asymmetric QW is
exposed to an in-plane magnetic field $H_y$ directed along the $y$
axis, the electron energy spectrum of the QW consists of a set of
subbands which are shifted along the $k_x$ axis with respect to
each other by the wave vector $\Delta k_x\propto H_y$
(\onlinecite{Ando_82}). This shifting, which is schematically
pictured in Fig.~1(b), leads to the asymmetric energy spectrum of
electrons, $\varepsilon(k_x)\neq\varepsilon(-k_x)$. Let an
electron system with such an asymmetric energy spectrum be
subjected to an electromagnetic field with the frequency
$\omega_0$. Then the resonant points of the intersubband
electron-photon interaction are positioned asymmetrically in the
$\mathbf{k}$-space with respect to subband edges. Correspondingly,
it is reasonable to expect that the photon-induced energy gaps
$\Delta\varepsilon$ will be positioned asymmetrically within the
subbands. The remarkable feature of such an asymmetrically--gapped
energy spectrum is the nondissipative flowing of electron gas. For
instance, let an electron gas fill dressed states under the Fermi
level $\mu$ [see Fig.~1(b)]. It is easy to show that the electric
current along the $x$ axis, produced by the electron gas, is
$j_x\propto\Delta\varepsilon$. Since this nonzero current is
associated to the ground state of the electron system, it flows
without dissipation. Thus, the photon-dressed electron system with
broken time-reversal symmetry can demonstrate the
superconductor-like behavior. The given paper is devoted to the
theoretical justification of this phenomenon.

The paper is organized as follows. In the Sec. II we  introduce
the electron-photon Hamiltonian and find its exact eigenstates and
eigenvalues. Section III is devoted to an analysis of electron
transport in photon-dressed QWs. Section IV contains a discussion
and conclusions.

\section{Electron-photon Hamiltonian} The electron energy
spectrum of a QW exposed to an in-plane magnetic field $H_y$ can
be described by the expression \cite{Ando_82}
\begin{equation}\label{En}
\varepsilon_n(\mathbf{k})=\varepsilon_{n0}-
\frac{\hbar\bar{z}_neH_y}{cm^\ast}k_x
+\frac{\hbar^2(k_x^2+k_y^2)}{2m^\ast},
\end{equation}
which takes into account linear magnetic-field terms. Here
$\mathbf{k}$ is the in-plane electron wave vector, $n=1,2,3...,$
is the number of electron subbands, $m^\ast$ is the effective
electron mass in the QW, $e$ is the electron charge,
$\varepsilon_{n0}$ is the energy of subband edge, and
$\bar{z}_n=\langle\psi_n(z)|z|\psi_n(z)\rangle$ is the averaged
$z$ coordinate of the electrons in the QW. Correspondingly,
electron wave functions in the subbands (\ref{En}) are
$\psi_n(\mathbf{k})=\exp{(i\mathbf{k}\mathbf{r})}\psi_n(z)$, where
$\mathbf{r}$ is the in-plane radius--vector, and the wave function
$\psi_n(z)$, arising from the confining potential $U(z)$, meets
the Schr\"odinger equation
$$-\frac{\hbar^2}{2m^\ast}\frac{d^2\psi_n(z)}{dz^2}+
U(z)\psi_n(z)=\varepsilon_{n0}\psi_n(z).$$ Let us restrict our
consideration by electron processes in the two lowest subbands
(\ref{En}) with $n=1,2$. Hereafter we will mark these subbands
$\varepsilon_{1,2}(\mathbf{k})$ and their wave functions
$\psi_{1,2}(\mathbf{k})$ by the symbols
$\varepsilon^\pm(\mathbf{k})$ and $\psi^\pm(\mathbf{k})$,
respectively, where the sign ``$-$'' corresponds to the lower
subband in Fig.~1b ($n=1$), and the sign ``$+$'' corresponds to
the upper one ($n=2$). Correspondingly, the shifting of the
electron subbands in $\mathbf{k}$--space, pictured in Fig.~1(b),
is $\Delta k_x=(\bar{z}_1-\bar{z}_2)eH_y/c\hbar$.

Let the QW be exposed to a plane monochromatic electromagnetic
wave with the frequency $\omega_0$ (dressing field), which is
linearly polarized along the $z$ axis. For simplicity, we will
neglect any spatial inhomogeneity of the dressing field but a
proper generalization can be easily made. To describe the
wave-induced mixing of the electron states from the two subbands
$\varepsilon^+(\mathbf{k})$ and $\varepsilon^-(\mathbf{k})$, the
electron Hamiltonian should be written as a $2\times2$ matrix
$\hat{\cal
H}_e=[\varepsilon^+(\mathbf{k})+\varepsilon^-(\mathbf{k})]\hat{I}/2+
[\varepsilon^+(\mathbf{k})-\varepsilon^-(\mathbf{k})]\hat{\sigma}_z/2$,
where $\hat{I}$ is the unity matrix, and $\hat{\sigma}_{x,y,z}$
are the Pauli matrices written in the basis of the two electron
states $\psi^\pm(\mathbf{k})$. As to the Hamiltonian of the
intersubband electron-wave interaction, it can be expressed within
the dipole approximation by $\hat{\cal
H}_\mathrm{int}=-{{{d}}}{E}\hat{\sigma}_x$, where
$\mathbf{E}=(0,\,0,\,E)$ is the electric field vector of the wave
(the dressing field vector), and
$d=e\langle\psi_1(z)|z|\psi_2(z)\rangle$ is the intersubband
matrix element of the electric dipole moment which is assumed to
be real and positive. Considering the problem within the
conventional quantum-field approach,
\cite{Cohen-Tannoudji_b98,Scully_b01} the classical field ${E}$
should be replaced with the field operator
$\hat{{E}}=i\sqrt{2\pi\hbar\omega_0/V}\left(\hat{a}-
\hat{a}^\dagger\right)$, where $V$ is the quantization volume of
the field, $\hat{a}$ and $\hat{a}^\dagger$ are the photon
operators of annihilation and creation, respectively, written in
the Schr\"{o}dinger picture (the representation of occupation
numbers \cite{Landau_4}). After such a replacement, the
interaction Hamiltonian takes the form $\hat{\cal
H}_\mathrm{int}=-i{d}\sqrt{{2\pi\hbar\omega_0}/{V}}\left(\hat{\sigma}_+\hat{a}+
\hat{\sigma}_-\hat{a}-\hat{\sigma}_-\hat{a}^\dagger-\hat{\sigma}_+\hat{a}^\dagger\right)$,
where $\hat{\sigma}_{\pm}=(\hat{\sigma}_x\pm i\hat{\sigma}_y)/2$.
In what follows, we will assume that the photon energy
$\hbar\omega_0$ is much more than the characteristic energy of the
electron-field interaction, $dE_0$, where $E_0$ is the field
amplitude. Then the terms $\hat{\sigma}_{+}\hat{a}^\dagger$ and
$\hat{\sigma}_{-}\hat{a}$ can be neglected, which corresponds to
the rotating-wave approximation commonly used in quantum optics
\cite{Cohen-Tannoudji_b98,Scully_b01}. As a result, the
interaction Hamiltonian is $\hat{\cal
H}_\mathrm{int}=-i{d}\sqrt{{2\pi\hbar\omega_0}/{V}}\left(\hat{\sigma}_+\hat{a}-
\hat{\sigma}_-\hat{a}^\dagger\right)$. Since the operator of the
field energy is $\hat{{\cal
H}}_0=\hbar\omega_0\hat{a}^\dagger\hat{a}$, the full Hamiltonian
of the considered electron-photon system, $\hat{{\cal
H}}=\hat{{\cal H}}_0+\hat{\cal H}_e+\hat{\cal H}_\mathrm{int}$,
reads as
\begin{eqnarray}\label{H}
\hat{{\cal
H}}&=&\hbar\omega_0\hat{a}^\dagger\hat{a}+\frac{\varepsilon^+(\mathbf{k})+\varepsilon^-(\mathbf{k})}{2}\hat{I}+
\frac{\varepsilon^+(\mathbf{k})-\varepsilon^-(\mathbf{k})}{2}\hat{\sigma}_z\nonumber\\
&-&i{d}\sqrt{\frac{2\pi\hbar\omega_0}{V}}\left(\hat{\sigma}_+\hat{a}-
\hat{\sigma}_-\hat{a}^\dagger\right).
\end{eqnarray}

The Hamiltonian (\ref{H}) is formally similar to the Hamiltonian
of the exactly solvable Jaynes-Cummings model. \cite{Jaynes_63}
Therefore, the Schr\"odinger problem with the electron-photon
Hamiltonian (\ref{H}) can also be solved exactly. Applying the
methodology \cite{Kibis_10,Kibis_11} to the problem, let us
introduce the joined electron-photon space
$|\pm,N\rangle=|\psi^\pm(\mathbf{k})\rangle\otimes|N\rangle$ to
describe the electron being in the state with the wave function
$\psi^\pm(\mathbf{k})$ and the dressing field being in the state
with the photon occupation number $N=1,2,3,...$. The basic states
of this space $|\pm,N\rangle$ are orthonormal and meet the
conditions $\langle\pm,N|\pm,N^\prime\rangle=\delta_{N,N^\prime}$
and $\langle\pm,N|\mp,N^\prime\rangle=0$. Therefore, the exact
eigenstates of the Hamiltonian (\ref{H}),
$\varphi_N^+(\mathbf{k})$ and $\varphi_N^-(\mathbf{k})$, can be
written as
\begin{eqnarray}\label{phi}
|\varphi_N^\pm(\mathbf{k})\rangle&=&\sqrt{\frac{\Omega_\pm(\mathbf{k})+|\omega(\mathbf{k})|}{2\,\Omega_\pm(\mathbf{k})}}
\,|\pm,N\rangle\nonumber\\
&+&i\eta(\mathbf{k})\sqrt{\frac{\Omega_\pm(\mathbf{k})-
|\omega(\mathbf{k})|}{2\,\Omega_\pm(\mathbf{k})}}\,|\mp,N\pm1\rangle,
\end{eqnarray}
where $\Omega_\pm(\mathbf{k})=\sqrt{8{d}^2(N+1/2\pm1/2)(\pi
\omega_0/\hbar V)+\omega^2(\mathbf{k})}$,
$\omega(\mathbf{k})=\omega_0-[\varepsilon^+(\mathbf{k})-\varepsilon^-(\mathbf{k})]/\hbar$,
and
\begin{equation}\label{eta}\nonumber
\eta(\mathbf{k})=\left\{\begin{array}{rl} -1,
&\omega(\mathbf{k})>0\\\\
1, &\omega(\mathbf{k})\leq0
\end{array}\right..
\end{equation}
The two energy branches corresponding to the electron-photon
states (\ref{phi}), $\varepsilon_N^+(\mathbf{k})$ and
$\varepsilon_N^-(\mathbf{k})$, are given by
\begin{equation}\label{Energy}
\varepsilon_N^\pm(\mathbf{k})=N\hbar\omega_0+\frac{\varepsilon^+(\mathbf{k})+\varepsilon^-(\mathbf{k})}{2}
\pm\frac{\hbar\omega_0}{2}\pm\eta(\mathbf{k})\frac{\hbar\Omega_\pm(\mathbf{k})}{2}.
\end{equation}
The expressions (\ref{phi}) and (\ref{Energy}) can be easily
verified by direct substitution into the Schr\"odinger equation
$\hat{{\cal
H}}\varphi_N^\pm(\mathbf{k})=\varepsilon_N^\pm(\mathbf{k})\varphi_N^\pm(\mathbf{k})$
with the Hamiltonian (\ref{H}), keeping in mind the relations
\cite{Landau_4,Landau_3}
\begin{eqnarray}
\hat{\sigma}_{\pm}\,|\mp,N\rangle&=&|\pm,N\rangle\,,\,\,\,\,
\hat{\sigma}_{\pm}\,|\pm,N\rangle\,=0\,,\,\,\,\nonumber\\
\hat{a}\,|\pm,N\rangle&=&\sqrt{N}\,|\pm,N-1\rangle,\,\,\,\,
\hat{\sigma}_{z}\,|\pm,N\rangle=\pm|\pm,N\rangle,
\nonumber\\
\hat{a}^{\dagger}\,|\pm,N\rangle&=&\sqrt{N+1}\,|\pm,N+1\rangle.\nonumber
\end{eqnarray}
As expected, the state $|\varphi_N^\pm(\mathbf{k})\rangle$ turns
into the state $|\pm,N\rangle$ when the electron-photon
interaction vanishes (i.e. when ${d}=0$). In the most interesting
case of a laser-generated intense dressing field (when the values
of $N$ and $V$ tend to infinity while the ratio $N/V$ is constant
\cite{Cohen-Tannoudji_b98}), the full energy of the
electron-photon system (\ref{Energy}) can be written as a sum
$N\hbar\omega_0+\varepsilon(\mathbf{k})$, where $N\hbar\omega_0$
is the energy of the dressing field, and
\begin{equation}\label{E}
\varepsilon(\mathbf{k})=\frac{\varepsilon^+(\mathbf{k})+\varepsilon^-(\mathbf{k})}{2}
\pm\frac{\hbar\omega_0}{2}\pm{\eta(\mathbf{k})}\frac{\Omega(\mathbf{k})}{2}
\end{equation}
is the desired energy spectrum of the dressed electrons. Here
$\Omega(\mathbf{k})=\sqrt{(\hbar\Omega_R)^2+[\hbar\omega_0-\varepsilon^+(\mathbf{k})+\varepsilon^-(\mathbf{k})]^2}$,
$\Omega_R=dE_0/\hbar$ is the Rabi frequency of intersubband
electron transitions, and $E_0=\sqrt{8\pi N\hbar\omega_0/V}$ is
the classical amplitude of the dressing field. The two branches of
the spectrum (\ref{E}) describe two subbands of dressed electrons,
which are pictured in Fig.~1(b) with solid lines: The branch with
the sign ``$+$'' corresponds to the upper pictured subband and the
branch with the sign ``$-$'' corresponds to the lowest one. As
expected, the energy gaps
\begin{equation}\label{G}
\Delta\varepsilon=\hbar\Omega_R,
\end{equation}
pictured in Fig.~1(b), are opened at the electron wave vector
$\mathbf{k}=\mathbf{k}_0$ satisfying the resonance condition,
$\hbar\omega_0=\varepsilon^+(\mathbf{\mathbf{k}_0})-\varepsilon^-(\mathbf{\mathbf{k}_0})$.
If the dressing field is absent ($E_0=0$), the expression
(\ref{E}) turns into the the spectrum of the free electron,
$\varepsilon_{1,2}(\mathbf{k})$, given by Eq.~(\ref{En}) and
pictured in Fig.~1(b) with thin lines. It should be noted that the
energy spectrum of dressed electrons, written in the form
(\ref{E}), is of universal character and applicable to describe
dressed electron states arising from any two photon-mixed energy
bands of free electrons, $\varepsilon^-(\mathbf{k})$ and
$\varepsilon^+(\mathbf{k})$, in any solid. Particularly, in the
case of a direct gap semiconductor, the energy spectrum of the
free electrons has the form
$\varepsilon^\pm(\mathbf{k})=\pm\varepsilon_g/2\pm\hbar^2\mathbf{k}^2/2m^\ast$,
where $\varepsilon_g$ is the semiconductor bandgap. Substituting
this expression into Eq.~(\ref{E}), we arrive at the energy
spectrum of dressed electrons in a bulk semiconductor, which is
pictured in Fig.~1(a) and was shown for the first time in
Ref.~\onlinecite{Elesin_69}.

\section{Transport properties of dressed electrons}
The electric current along the $x$ axis, which is produced by a
dressed electron with the wave vector $\mathbf{k}$, is
$j_x({\mathbf{k}})=ev_x({\mathbf{k}})/L_x$, where $L_{x,y}$ are
the in-plane dimensions of the QW along the $x,y$ axis, and
$\mathbf{v}({\mathbf{k}})$ is the average velocity of the dressed
electron in the state (\ref{phi}). This average velocity is given
by the conventional quantum-mechanical expression
$\mathbf{v}({\mathbf{k}})=
\langle\varphi_N^\pm(\mathbf{k})|\hat{\mathbf{v}}|\varphi_N^\pm(\mathbf{k})\rangle$,
where $\hat{\mathbf{v}}=(i/\hbar)[\hat{{\cal
H}}\hat{\mathbf{r}}-\hat{\mathbf{r}}\hat{{\cal H}}]$ is the
operator of the electron velocity, and $\hat{\mathbf{r}}$ is the
operator of the electron coordinate. To calculate the matrix
element
$\langle\varphi_N^\pm(\mathbf{k})|\hat{\mathbf{v}}|\varphi_N^\pm(\mathbf{k})\rangle$,
we will use the $\mathbf{k}$--representation of the coordinate
operator, $\hat{\mathbf{r}}=i\partial/\partial\mathbf{k}$
(\onlinecite{Landau_3}). Then the expression for the velocity of
dressed electron takes the form
$\mathbf{v}(\mathbf{k})=(1/\hbar)\partial\varepsilon^{\pm}_N(\mathbf{k})/\partial\mathbf{k}$.
As expected, this expression coincides formally with the
well-known classical Hamilton equation,
$\mathbf{v}(\mathbf{p})=\partial\varepsilon(\mathbf{p})/\partial\mathbf{p}$,
describing the velocity of a particle with an energy
$\varepsilon(\mathbf{p})$ and the generalized particle momentum
$\mathbf{p}=\hbar\mathbf{k}$. In the case of the many-electron
system, the full current along the $x$ axis is
$j_x=\sum_{\mathbf{k}}j_x({\mathbf{k}})$, where the summation
should be performed over filled states (\ref{phi}). Taking into
account the aforesaid, the current produced by dressed electrons
lying under the Fermi level $\mu$ [see Fig.~1(b)] is
$j_{x0}=(eL_y/2\pi^2\hbar
)\int(\partial\varepsilon(\mathbf{k})/\partial k_x)d^2\mathbf{k}$,
where the integration should be performed over the spectrum
(\ref{E}) with the energy $\varepsilon(\mathbf{k})\leq\mu$. As a
result of the integration, we arrive at the expression
\begin{equation}\label{J}
j_{x0}=\frac{eL_y}{\pi^2\hbar^2}\sqrt{{2m^\ast\mu}}\Delta\varepsilon.
\end{equation}

Since the current (\ref{J}) is associated to the ground state of
the photon-dressed electron system, it flows without dissipation
and should be considered as a persistent current. It follows from
Eq.~(\ref{J}) that the current arises from the photon-induced
energy gap (\ref{G}): If the dressing field is absent ($E_0=0$),
both the gap (\ref{G}) and the current (\ref{J}) vanish. The
physical reason of the interrelationship between the persistent
current (\ref{J}) and the gap (\ref{G}) is clarified in Fig.~2(a).
Since the gap (\ref{G}) forbids an elastic backscattering of
electrons from states lying in the energy range
$\Delta\varepsilon$, a current associated to these electron states
flows without dissipation. Therefore, we marked these states in
Fig.~2(a) as ``electron states with persistent current''. As
expected, the full current of the marked states is exactly equal
to the persistent current (\ref{J}). It should be stressed that
the gap (\ref{G}) arises from stationary solutions of the
time-independent Schr\"odinger problem with the stationary
Hamiltonian (\ref{H}) describing the closed system ``electron +
quantized electromagnetic field''. Therefore, it is the true
(stationary) gap in the density of electron-photon states
(\ref{phi}). As a consequence, the gap (\ref{G}) will manifest
itself directly in all phenomena sensitive to the density of
states of charge carriers, including backscattering processes
marked by the arrow in Fig.~2(a).

It follows from the charge conservation law that an electric
current in any conductor must satisfy the continuity condition.
Applying this general rule to the considered system, we arrive at
the evident result: The ground state of a QW with the nonzero
current (\ref{J}) satisfies the continuity condition only if the
QW is a part of a closed electrical circuit, where the current
$j_x=j_{x0}$ flows [see Fig.~2(b)]. If the circuit is broken, the
continuity condition requires the zero current, $j_x=0$.
Particularly, a state of the photon-dressed electron system with
the zero current always takes place in an isolated QW: Though the
state is not ground, it corresponds to the minimal energy under
the additional condition of $j_x=0$. As a consequence, an expected
voltage-current characteristic of the QW, $V(j_x)$, takes the form
pictured schematically in Fig.~2(c). The remarkable feature of the
characteristic is the zero electrical resistance in the broad
range of currents, $j_{x0}<j_x<0$, which arises from the electron
states with persistent current. Though such a zero-resistance
behavior of photon-dressed QW is superconductor-like, the physical
reason of the declared effect differs conceptually from both the
conventional superconductivity in solids and superfluidity in
quantum liquids. Indeed, these known mechanisms of the
nondissipative flow are based on interaction processes in strongly
correlated many-particle systems, whereas the discussed phenomenon
arises from the one-electron Hamiltonian (\ref{H}) and takes place
for a photon-dressed gas of noninteracting electrons. Thus, the
declared effect results in an unconventional superconductivity
which is discussed below.
\begin{figure}[th]
\includegraphics[width=0.48\textwidth]{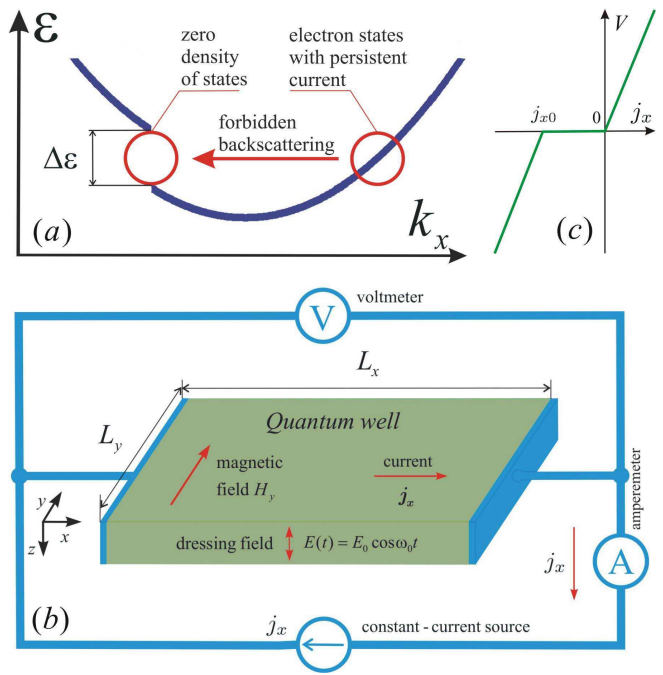}
\caption{(Color online) Scheme of the light-induced
superconductivity: (a) asymmetrically--gapped energy spectrum of
electrons as a physical reason of the persistent current; (b)
sketch of a proposed experimental setup for electron transport
measurements in a photon-dressed quantum well; (c) voltage-current
characteristic of the photon-dressed quantum well.}
\end{figure}

\section{Discussion and conclusions}
The electron-photon Hamiltonian (\ref{H}) does not take into
account a scattering of electrons, which leads to the finite
lifetime of electron states (\ref{En}), $\tau$. Therefore, the
results obtained above from the Hamiltonian (\ref{H}) are correct
if the photon-induced energy gap $\Delta\varepsilon$ is much large
than the scattering-induced washing of the electron energy
spectrum, $\hbar/\tau$. Taking into account Eq.~(\ref{G}), this
condition of applicability of the Hamiltonian (\ref{H}) can be
written in the form
\begin{equation}\label{R}
\Omega_R\tau\gg1,
\end{equation}
which coincides with the condition of photon-induced Rabi
oscillations of electrons between the subbands
$\varepsilon^-(\mathbf{k})$ and $\varepsilon^+(\mathbf{k})$
(\onlinecite{Scully_b01}). Thus, the gap (\ref{G}) --- and,
correspondingly, the light-induced persistent current (\ref{J})
--- takes place when the electron subsystem is in the regime of
intersubband Rabi oscillations. The coexistence of the persistent
current and the Rabi oscillations has a deep physical meaning. To
clarify it, let us write the eigenstates (\ref{phi}) in the
resonant point $\mathbf{k}=\mathbf{k}_0$, where the gap (\ref{G})
is opened. Since the wave vector $\mathbf{k}_0$ satisfies the
condition $\omega(\mathbf{k}_0)=0$, the eigenstates (\ref{phi}) at
$\mathbf{k}=\mathbf{k}_0$ are
\begin{equation}\label{k0}
|\varphi_N^\pm(\mathbf{k}_0)\rangle=
\frac{1}{\sqrt{2}}\left[|\pm,N\rangle+
i\,|\mp,N\pm1\rangle\right],
\end{equation}
and their energies (\ref{E}) are
$\varepsilon^\pm(\mathbf{k}_0)\pm\Delta\varepsilon/2$,
respectively. It is seen that the bound electron-photon states
(\ref{k0}) describe the one-photon mixing of the electron subbands
$\varepsilon^-(\mathbf{k})$ and $\varepsilon^+(\mathbf{k})$, which
corresponds physically to the Rabi oscillations of the electron
subsystem between these two subbands. As to the gap (\ref{G}), it
can be treated as a total binding energy of the two coherent
electron-photon states (\ref{k0}) with the two different signs
``$\pm$''. Thus, the existence of the photon-induced coherent
states (\ref{k0}) with the binding energy (\ref{G}) is
microscopical reason of the discussed effect.

It should be noted that an electron, performing periodical Rabi
oscillations between the subbands $\varepsilon^-(\mathbf{k})$ and
$\varepsilon^+(\mathbf{k})$, does not absorb the energy of
electromagnetic field inducing these Rabi oscillations. In other
words, the energy of the dressing field averaged over the period
of Rabi oscillations is constant. As a consequence, the condition
of Rabi oscillations (\ref{R}) is physically equal to the
forbidding of all processes accompanied with the absorption of the
field energy by electrons. Therefore, the inequality (\ref{R}) can
be considered as a condition of a purely dressing (nonabsorbable)
electromagnetic field. As a result, the existence of a persistent
current under the condition (\ref{R}) does not contradict the
energy conservation law, since the nondissipative flow of
photon-dressed electrons is not accompanied with the absorbing of
energy of the dressing field.

It follows from the aforesaid that the light-induced persistent
current (\ref{J}) differs conceptually from light-induced currents
which arise from a photovoltaic effect. Indeed, any photovoltaic
effect must be accompanied with absorbing light energy by
electrons (see, e.g., Ref. \onlinecite{Sturman_92}) and,
therefore, it is absent under the condition (\ref{R}). On the
contrary, if the condition (\ref{R}) is broken, the intersubband
absorption of light
--- and, correspondingly, the photovoltaic effect
--- appears. In other words, there are two different regimes of
electron-photon interaction in QWs with broken time-reversal
symmetry: (I) The regime of weak electron-photon coupling
($\Omega_R\tau\alt1$), where an usual ohmic current appears from
the photovoltaic effect, and (II) the regime of strong
electron-photon coupling ($\Omega_R\tau\gg1$), where the
persistent current (\ref{J}) exists. The first regime has been
studied both theoretically and experimentally
\cite{Gorbatsevich_93,Aleshchenko_93,Omelyanovskii_96,Diehl_07,Diehl_09},
whereas the second one escaped attention before. It should be
stressed that the light-induced ohmic current and the
light-induced persistent current, which arise from the two
different physical mechanisms, flow in mutually opposite
directions. Therefore, they can be easily differentiated in
experiments.

Since the synthesis of semiconductor QWs with such characteristic
parameters as an energy interval between electron subbands
$[\varepsilon^+(\mathbf{k}_0)-\varepsilon^-(\mathbf{k}_0)]\sim10^{-1}$
eV, electron lifetime $\tau\sim10^{-10}$ s, intersubband dipole
moment $d\sim10^{2}$ D, and Fermi energy $\mu\sim10^{-2}$ eV is
routine procedure for the modern nanotechnology, the appropriate
source of a dressing field for the experimental observation of the
declared effect seems an infrared laser --- for instance, an
ordinary CO$_2$-laser with the wavelength $\lambda=10.6\,\,\mu$m
and the output power $P\agt10$ W/cm$^2$. It should be noted that
infrared lasers have been discussed as prospective sources of a
dressing field for observing dressed electron states in
semiconductors \cite{Mizumoto_06}. In our case, such a laser
satisfies both the condition of the resonant electron-photon
interaction,
$\hbar\omega_0=\varepsilon^+(\mathbf{k}_0)-\varepsilon^-(\mathbf{k}_0)$,
and the condition (\ref{R}). The current (\ref{J}) induced by this
laser field can be estimated for the above-mentioned parameters as
$j_{x0}/L_y\agt10^{-2}$ A/m. Thus, the persistent current
(\ref{J}) is large enough to be observed experimentally in actual
semiconductor nanostructures.

Though we calculated the persistent current (\ref{J}) for the
electron distribution corresponding to zero temperature, the
discussed effect can also exist at nonzero temperatures. Indeed,
the asymmetrically--gapped energy spectrum pictured in Fig.~2(a)
leads always to a nonzero current if an electron distribution
function depends only on electron energy. The particular case of
such a distribution function is the Fermi-Dirac function
describing an electron distribution on dressed states (\ref{E}) in
the thermodynamic equilibrium at any temperature. Since there is
no dissipation processes in the thermodynamic equilibrium, the
discussed persistent current takes place at nonzero temperatures
as well. In other words, the discussed superconductivity can be
high--temperature: It exists unless the increasing of electron
scattering --- which always takes place with increasing
temperature --- washes the gap (\ref{G}).

Generally, the light-induced persistent current needs broken
time-reversal symmetry. In the current paper we have considered an
electron system which is devoid of time-reversal symmetry due to a
magnetic field. However, the similar effect can also take place
for a time-reversally symmetric electron system interacting with a
dressing field without time-reversal symmetry. For instance, a
circularly polarized field is devoid of time-reversal symmetry,
since the time reversal turns clockwise polarized photons into
counterclockwise polarized ones and vice versa. Therefore, the
electron coupling to a circularly polarized field can result in
the persistent current in various time-reversally symmetric
electron systems --- particularly, in curvilinear quantum wires
\cite{Kibis_11}. However, a fabrication of the quantum wires is
not a trivial technological problem, that impedes an experimental
observation of the discussed effect in one-dimensional conductors.
On the contrary, semiconductor QWs with a high-mobility
two-dimensional electron gas can be easily synthesized in a modern
laboratory. As a result, the significant new area of experimental
research in QWs
--- where the physics of superconductivity, the physics of
nanostructures, and quantum optics meet --- can be opened.

It should be noted that there is another kind of persistent
current associated with the ground state of an electron system
with broken time-reversal symmetry. This known persistent current
arises from the Aharonov-Bohm effect \cite{Aharonov_59} and takes
place in quantum rings exposed to a magnetic field
\cite{Buttiker_83,Mailly_93}. However, the Aharonov-Bohm
persistent current cannot be identified with superconductivity,
since this current is closed. Indeed, it flows in a microscopical
ring and is devoid of such a characteristic property of
superconductivity as a dissipationless carrying of electrons over
a macroscopically long distance. In contrast to the Aharonov-Bohm
persistent current, the light-induced persistent current (\ref{J})
flows in a QW with macroscopically large in-plane dimensions
$L_{x,y}$ [see Fig.~2(b)]. Therefore, the persistent current
(\ref{J}) leads to the dissipationless carrying of an electric
charge over the macroscopically long distance $L_{x}$. This allows
to consider the discussed effect as a conceptually novel mechanism
of superconductivity.

Finalizing the discussion, it should be noted that the appearance
of unusual effects of strong electron-photon coupling is common
feature of quantum systems with broken fundamental symmetries,
including both the broken time-reversal symmetry \cite{Kibis_11}
and the broken inversion symmetry \cite{Kibis_09}. In the current
paper we have solved the spinless problem, where the considered
effect arises from a diamagnetic transformation of the electron
energy spectrum (\ref{En}) by a magnetic field $H_y$. However, the
similar effect can also appear in photon-dressed spin systems
without time-reversal symmetry. Since an analysis of
spin-originated effects goes beyond the scope of the current
paper, it will be done elsewhere.

In summary, we have declared the nondissipative flowing of
photon-dressed electron gas (the light-induced superconductivity),
which differs conceptually from both the superfluidity of quantum
liquids and conventional superconductivity in solids. This
macroscopic quantum phenomenon lies in the scientific area, where
condensed-matter physics and quantum optics meet. It is of
universal character and can take place in various strongly coupled
electron-photon systems with broken time-reversal symmetry.
Particularly, the phenomenon can be observed in asymmetric quantum
wells exposed to an in-plane magnetic field. Since the fabrication
of such quantum wells is routine procedure for modern
nanotechnology, the declared phenomenon can take place in actual
nanostructures.
\begin{acknowledgements}
The work was partially supported by the RFBR (Project No.
10-02-00077), the Russian Ministry of Education and Science, and
the Seventh European Framework Programme (Project No. FP7-230778).
\end{acknowledgements}

\end{document}